\documentclass[preprint,12pt]{elsarticle}

\usepackage{amssymb}
\usepackage{graphicx}
\usepackage{dcolumn}
\usepackage{bm}
\usepackage[usenames]{color}

\usepackage[T1]{fontenc}
\usepackage{mathptmx}

\journal{Ceramics}

\begin{document}

\begin{frontmatter}



\title{Synthesis of Electron- and Hole-Doped Bulk $BaFe_2As_2$ Superconductors by Mechanical Alloying}


\author{K.S. Pervakov, V.A. Vlasenko}

\address{Ginzburg Center for High-Temperature Superconductivity and Quantum Materials, 
	P.N. Lebedev Physical Institute of the Russian Academy of Sciences (LPI RAS)
	Leninsky ave. 53, Moscow 119991, Russia}

\begin{abstract}
Here, we report the successful synthesis of electron- (Ni) and hole- (K) doped bulk $BaFe_2As_2$ compounds by means of mechanical alloying (MA) process. The MA technique allows one to obtain amorphous phase $BaFe_2As_2$ with a homogeneous distribution of doping atoms at room temperature without continuous heating. We found the optimum time for high-energy milling is in the range between 1-1.5 hours. Using pre-synthesised $FeAs$ and $NiAs$ precursors, we successfully obtained  $BaFe_{2-x}Ni_xAs_2$ and $Ba_{1-x}K_xFe_2As_2$ superconducting bulk samples after short-range heat treatment. This method can be easily scaled according to industrial needs to provide high-quality bulk superconducting material for wires, trapped-field magnets and other applications.

\end{abstract}


\end{frontmatter}


\section{INTRODUCTION}

Mechanical alloying (MA) is a widespread processing method implying grinding powder material in a high-energy ball mill. To date, MA has been used in synthesising a variety of equilibrium and non-equilibrium alloy phases from precursor powders \cite{Suryanarayana1}. The mechanical alloying method has been successfully used in the preparation of superconductors such as $Nb_3Al$ \cite{Yan}, $MgB_2$ \cite{Gao,Yudanto},  $DyBa_2Cu_3O_{6+\delta}$ \cite{Dy}, $Bi-Sr-Ca-Cu-O$ system \cite{Tessier}, $YBa_2Cu_3O_{7-d}$ \cite{Hamrita, Slimani}, $FeSe_{1-x}$ \cite{Ulbrich,Xia}, $SmFeAsO$ \cite{Shi}, $SmFe_{1-x}Co_{x}AsO$ \cite{Yang}, $SmFe_{1-x}Co_{x}AsO_{1-x}F_x$ \cite{Yang2}, $Ba_{0.6}K_{0.4}Fe_2As_2$ \cite{Li2, Weiss}. Therefore, mechanical alloying may become the main method to form a high-quality homogeneous precursor or parent superconducting compound. Among superconductors, iron-based superconductors (IBS) \cite{Kamihara} are still a point of interest due to their unusual physical properties and high critical parameters such as upper critical field ($H_{c2}$), critical current density ($J_c$), and low anisotropy. In the IBS class, the $BaFe_2As_2$ (122) superconducting family is the most promising candidate for practical application \cite{Hosono}. 
For the parent $BaFe_2As_2$, the superconductivity (SC) induces by doping on all lattice sites. Notably, electron-doping to FeAs-layers can be achieved by partial substituting the iron site with appropriate transition metals ($Co^{2+}, Ni^{2+}$) \cite{Luo, Xu1}. The hole-doping in the 122 phase is attained SC by partial substitution of $Ba^{2+}$ sites by $K^{1+}$ or $Na^{1+}$ etc \cite{Nakajima1, Luo2}. The SC state also can be obtained by isovalent atomic substitution of $As^{3+}$ for  $P^{3+}$ and $Fe^{4+}$ for $Ru^{4+}$ \cite{Eom}.
The upper critical field in $Ba-122$ superconducting compounds is estimated to be more than 60T, with a critical current of up to $10^6 A/cm^2$ \cite{Hosono,Pervakov,Shimoyama}. Thus, such superconductors are applicable for high-field and high-current application such as superconducting wires, magnets, fault current limiters etc. The industry needs bulk material in large quantities that cannot be produced by laboratory techniques such as synthesis in a quartz ampoule; so, we used the scalable MA method to manufacture $BaFe_{2-x}Ni_xAs_2$ (BFNA) and $Ba_{1-x}K_xFe_2As_2$ (BKFA) bulk superconductors. The sample quality was estimated by characterisation using powder X-ray diffraction, scanning electron microscope with an energy dispersive spectroscopy attachment and magnetic susceptibility measurements.

\section{METHODS}

\subsection{Mechanical Alloying Technique}
Mechanical activation or mechanical alloying is a well-known technique \cite{Carey, Avvakumov} for producing high homogeneity ceramic products from non-fibrous or non-elastic materials in various quantities. The MA process depends mainly on the milling jar volume, ball-to-powder weight ratio and initial size of the material. The process can be carried out under various ambient conditions such as vacuum, inert gas, oxidising or reducing atmosphere, reactive gas or liquid at high and low temperatures  \cite{Suryanarayana2}. The essence of MA is in repeated flattering, cold welding, fracturing and rewelding of the powder particles during high-energy milling. During the collision of the milling balls, some amount of powder is trapped and plastically deforms in between the balls. Thus, the initial powder particles tend to weld together and form large particles. In case the rate of defects accumulation exceeds the rate of defects relaxation, mechanical activation occurs. The following deformation creates conditions for phase formation without continuous heating of the components. However, after the mechanical alloying is processed, it is often necessary to carry out additional heat treatment of the homogeneous material to provide ordered crystal structure formation. Nevertheless, the heat treatment time is significantly reduced compared to classical solid-phase synthesis \cite{Rotter}.

In this work, we used a $Fritsch$ $Pulverisette$ $7$ $Premium$ $Line$ planetary ball mill to prepare the  $BaFe_{2-x}Ni_xAs_2$ (BFNA) and $Ba_{1-x}K_xFe_2As_2$ (BKFA) bulk samples. Mechanical alloying was carried out in a tungsten carbide milling garment with a volume of 45 ml with a batch of 5mm tungsten carbide balls. The volume ratio for tungsten balls, grinding material and free space was $\sim 1:1:1$, accordingly. We used metallic $Ba$, $K$ and pre-synthesized binary compounds $FeAs$ and $NiAs$ in a stoichiometric ratio as starting materials. The initial reagents were: $Ba$ (99,9$\%$) + $FeAs$ (99,9$\%$ $Fe$ + 99,9999$\%$ $As$) + $NiAs$ (99,95$\%$ $Ni$ + 99,9999$\%$ $As$) for electron-doped $BaFe_{2-x}Ni_xAs_2$ and $Ba$ (99,9$\%$) + $K$ (99,99$\%$) + $FeAs$ (99,9$\%$ $Fe$ + 99,9999$\%$ $As$) for hole-doped $Ba_{1-x}K_xFe_2As_2$. The required 122 compounds were prepared in several steps as follows. First, we weighed the reagents in a stoichiometric ratio according to the doping level and placed them into the milling jar together with tungsten carbide milling balls. After that, the volume of the milling jar was evacuated down to $10^{-3}$ Pa by a rotary pump with a $LN_2$ trap and placed into the planetary ball mill. The milling process was carried out 7-10 times at 900-1000 rpm for 5 minutes, followed by standing for 3 minutes. Short-term heat treatment for 1 hour was used to restore the long-range ordering of the crystal structure. The treated powder was examined by X-ray diffraction and magnetic susceptibility measurements. All manipulations were performed in a glove box with an argon atmosphere because of the extremely high oxidation of metallic Ba and K in the open air.

\subsection{Phase, Structure and Superconducting Properties Analysis}
The crystal structure and phase composition were studied using X-ray powder diffraction (XRD) before and after the MA process with a Rigaku MiniFlex 600 using $Cu-K_\alpha$ radiation in the 2$\Theta$ angle region from $10^o$ to $110^o$. The airtight sample holder with high purity ($>99.998\%$) argon atmosphere was used for XRD measurements to prevent oxidation of the air-sensitive components.  Content of moisture and oxygen in the argon was less than 0.1 ppm. Additional element analysis was done using a Scanning Electron Microscope (SEM) JEOL 7001F with a field-emission cathode and INCA X-Act Energy Dispersive Spectroscopy (EDS) attachment.

Superconducting properties of the bulk samples were studied by magnetic susceptibility measurements using a Quantum Design Physical Property Measurement System (PPMS-9) in the temperature range down to 2K. AC-susceptibility measurements were carried out in an AC field with an amplitude of 5 $Oe$ and a frequency of 337 Hz. The data were collected upon warming after cooling in a zero magnetic field. The M(H) measurements were done with vibrating sample magnetometer (VSM), and the sweep rate was 150 Oe/s.

\section{Results and discussion}

We successfully synthesised the electron-doped compound $BaFe_{2-x}Ni_xAs_2$ with doping levels of $x$=0.08, 0.10  and hole-doped $Ba_{1-x}K_xFe_2As_2$ samples with doping levels of $x$=0.35, 0.40. XRD measurements were carried out to study the phase formation process during the mechanical alloying. Fig. 1 shows the XRD patterns for the optimally doped BKFA compound depending on milling time. Our X-ray data clearly show that during the MA, the peaks related to the initial materials are suppressed and that the peaks associated with the emergence of the 122 phase appear. Similar behaviour for the nickel-doped samples was observed. The Rietveld phase analysis method (RIR) calculation data shown in Fig. 2 indicate that the main part of the 122 phase forms within 1 hour of MA time, followed by a slight increase in the amount of the 122 phase during MA up to 1.5 hours. Further grinding did not show any significant positive effect on the formation of the $BaFe_2As_2$ phase. Our results are similar to those presented in ref. \cite{Weiss}, where the temperature of the steel milling jar, which depends on the milling time, was used as a phase formation indicator. However, unlike to work mentioned above, we investigated the phase composition dependence on the MA time.

Mapping-mode SEM images for non-heat-treated bulks $BaFe_{1.9}Ni_{0.1}As_2$ and $Ba_{0.6}K_{0.4}Fe_2As_2$ are shown in Fig. 3 (a-d) and Fig. 3 (e-h), respectively. The non-uniform picture is caused by the powder form of the samples for SEM that cannot be solid as pressed ones with a smooth surface. That is why one can see shadows on the surface as dark spots imaging the topology of the surface. In both cases, the doping element - $Ni$/$K$, spreads rather evenly, which confirms a homogeneous dopant distribution. The quantitative composition of the investigated bulk samples in terms of atomic percentage is $Ba$-20.34$\%$, $Fe$-37.96$\%$, $Ni$-1.84 $\%$, $As$-39.86 $\%$ for $BaFe_{1.9}Ni_{0.1}As_2$ and $Ba$-15.13$\%$, $Fe$-39.51$\%$, $K$-5.61$\%$, $As$-39.75$\%$ for $Ba_{0.6}K_{0.4}Fe_2As_2$, which corresponds to a $Ba:Fe:Ni:As$ element ratio of \\ 1.022:1.907:0.093:2.003 and $Ba:K:Fe:As$ element ratio of 0.729:0.271:1.91:1.92. One can see the composition is in a good agreement with the ratio of the initial components used for milling. EDS data are presented in Table.1.

\begin{figure}[b]
	\includegraphics [width=1 \textwidth]{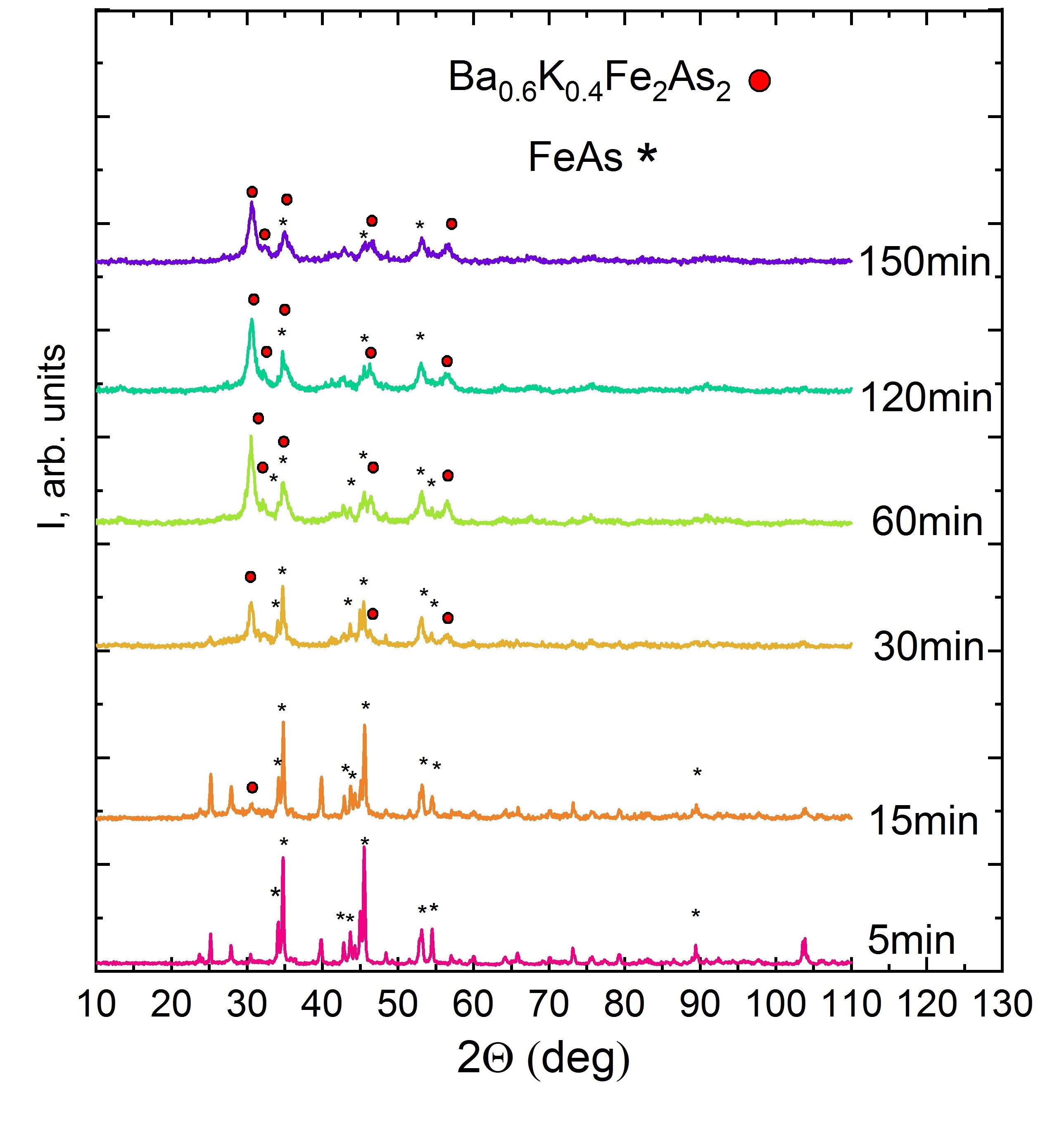}
	\caption{ XRD patterns of $Ba_{0.6}K_{0.4}Fe_2As_2 $ after different milling times.}	
\end{figure}

\begin{figure}[t]
	\includegraphics [width=1 \textwidth]{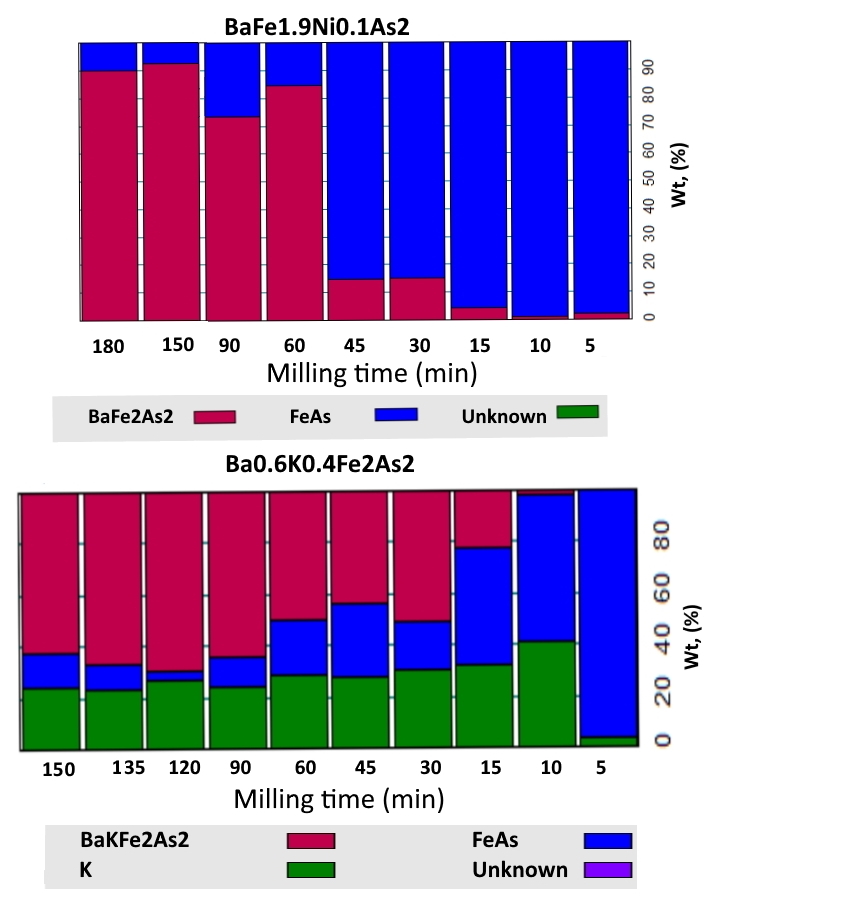}
	\caption{ Rietveld Quantitative Phase Analysis for bulk BKFA and BFNA depending on the MA time. }
	
\end{figure}

\begin{figure}[t]
	\includegraphics [width=1 \textwidth]{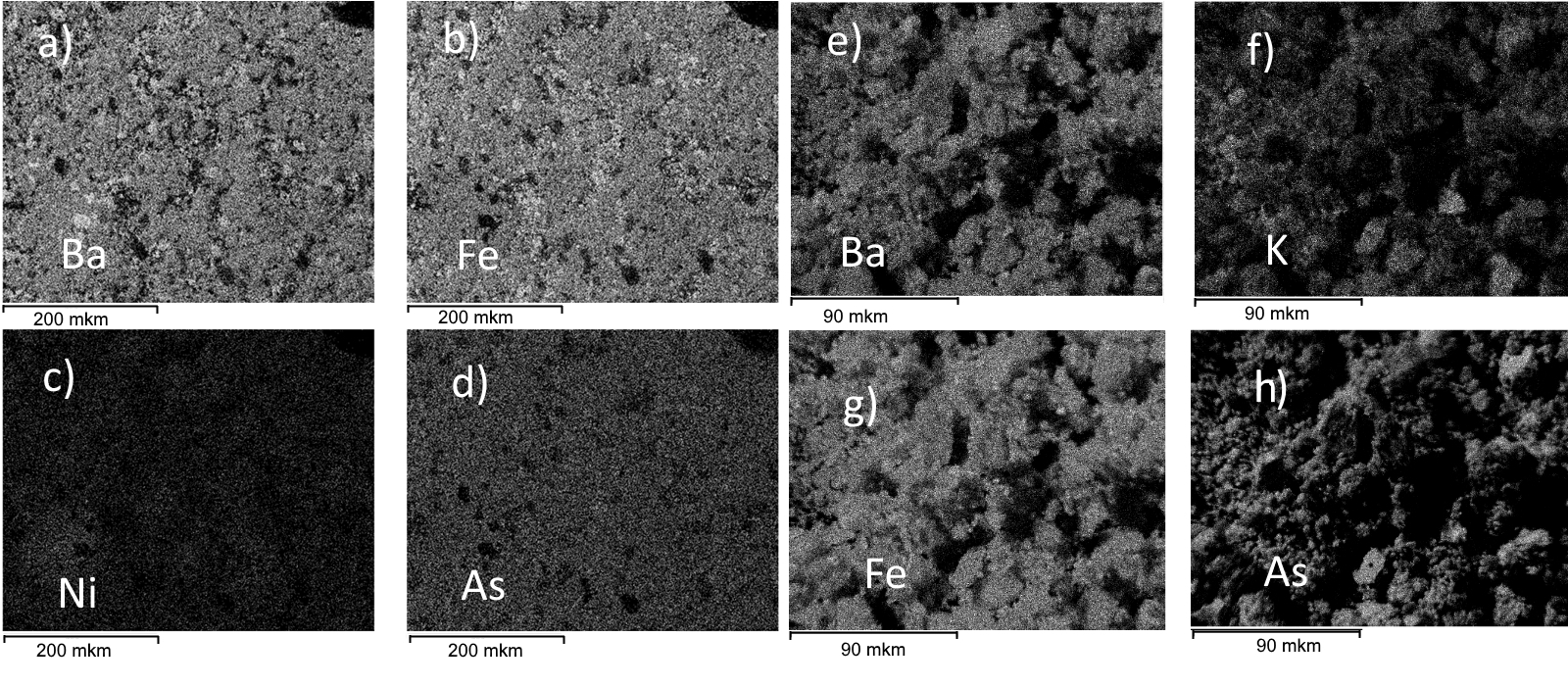}
	\caption{EDS mapping images of powder (a-d) $BaFe_{1.9}Ni_{0.1}As_2$ and (e-h) $Ba_{0.6}K_{0.4}Fe_2As_2$ samples. Dark spots in the mapping images are caused by the surface topology.}
\end{figure}

After grinding, we pressed the obtained BKFA powder into pellets and heat-treated them within 1 hour at temperatures ranging from 500$^0C$-950$^0C$ to determine the optimal preparation conditions. We provided magnetic susceptibility measurements after the MA process to investigate the formation of $Ba_{0.6}K_{0.4}Fe_2As_2$ and $BaFe_{1.9}Ni_{0.1}As_2$ superconducting phases. We found the obtained compounds did not exhibit superconductivity during the MA process. Thus, we suppose the formation of amorphous BKFA and BFNA compounds without long-range ordering or a $BaFe_2As_2$ parent compound with evenly distributed $K/Ni$ dopants. Superconductivity occurs only after heat treatment of the milled material. The optimal annealing temperature was estimated to be 850$^0C$. Pellets of BFNA were sintered for 1 hour at the same optimal temperature of 850$^0C$ determined for BKFA.

Structural changes after heat treatment were also examined by SEM and XRD. Fig. 4 shows the XRD patterns for near-optimum doped BKFA and BFNA superconducting samples after mechanical alloying within one-hour and following 1-hour heat treatment at 850$^0C$. The figure shows the domination of BFNA and BKFA superconducting phases, with only a small amount of unreacted compounds. From the X-ray diffraction data, we calculated the cell parameters for $Ba_{0.6}K_{0.4}Fe_2As_2$ ($a$=3.9405(7) \AA, $c$=13.195(3) \AA) and for $BaFe_{1.9}Ni_{0.1}As_2$ ($a$=3.9586(2) \AA, $c$=12.9820(7) \AA) bulk samples. The volume of the 122 phase in all cases was in the range 80-90$\%$, as determined from RIR.

The SEM images of the $BaFe_{1.9}Ni_{0.1}As_2$ pellet show the surface before and after heat treatment in Fig. 5(a) and Fig. 5(c), respectively. In the upper image (Fig. 5(a)), one can see small particles of the milled compound without oriented crystallites. One can observe the same picture for the $Ba_{0.6}K_{0.4}Fe_2As_2$ sample in Fig. 5(b, d). Therefore, annealing leads to the rapid growth of crystallites, which may indicate restoration of the long-range ordering in $BaFe_{1.9}Ni_{0.1}As_2$ and $Ba_{0.6}K_{0.4}Fe_2As_2$ compounds. The average size of the crystallites lies in a range of 3-5 $\mu$m. Crystallisation is also manifested in the XRD patterns as high-intensity narrow reflexes (see Fig. 4).

Thus, we show the formation of the amorphous Ba-122 compounds during mechanical alloying and the BKFA/BFNA superconducting phases after short heat treatment. The total time for synthesis is significantly reduced compared to the classical solid-phase reaction method \cite{Nikolo}.

\begin{figure}[t]
	\includegraphics [width=0.85 \textwidth]{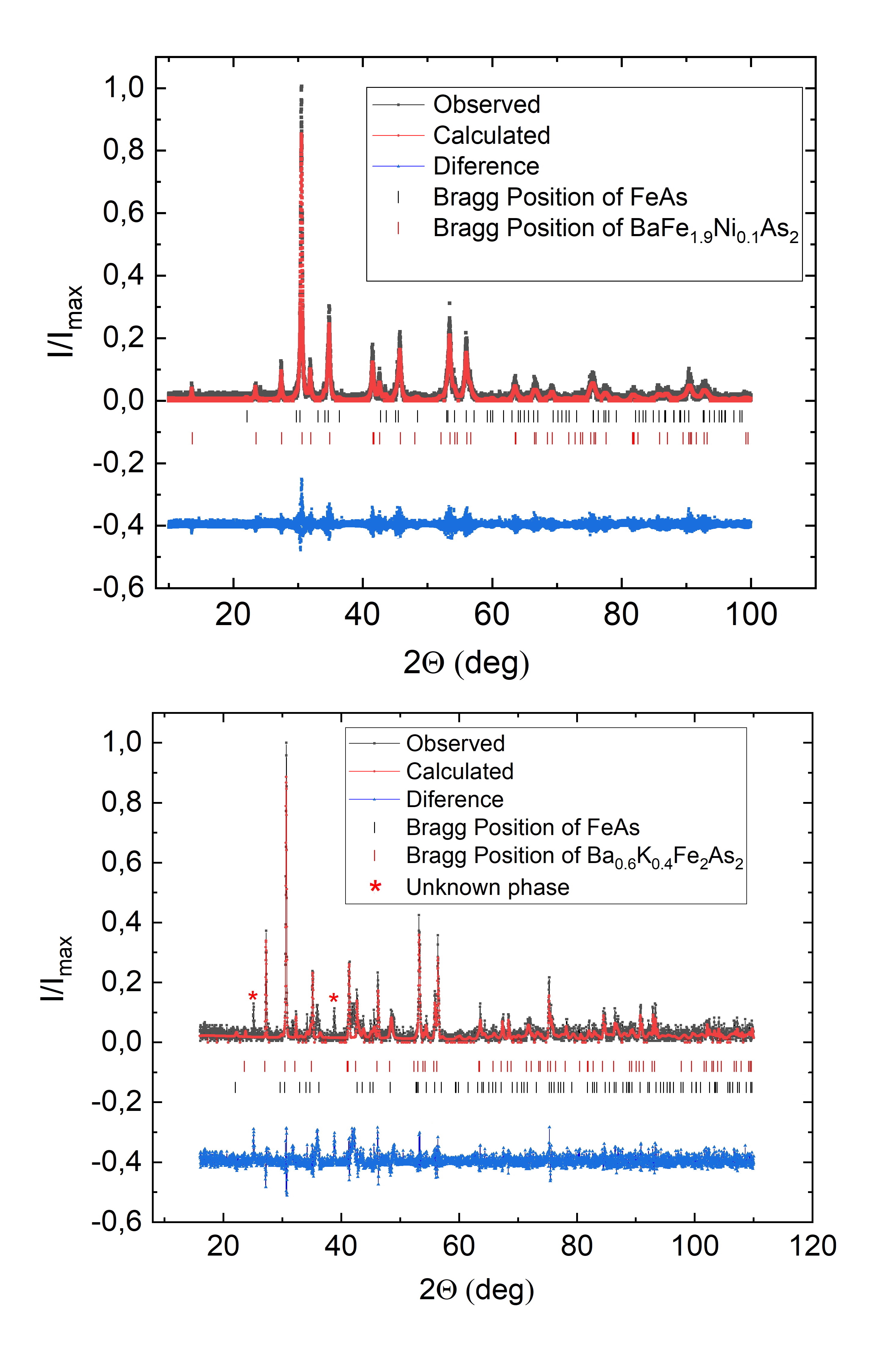}
	\caption{ Powder X-ray diffraction patterns and Rietveld refinement profile of the $Ba_{0.6}K_{0.4}Fe_2As_2 $ and $BaFe_{1.9}Ni_{0.1}As_2$ compounds after annealing.} 
	
\end{figure}

\begin{figure}[t]
	\includegraphics [width=1 \textwidth]{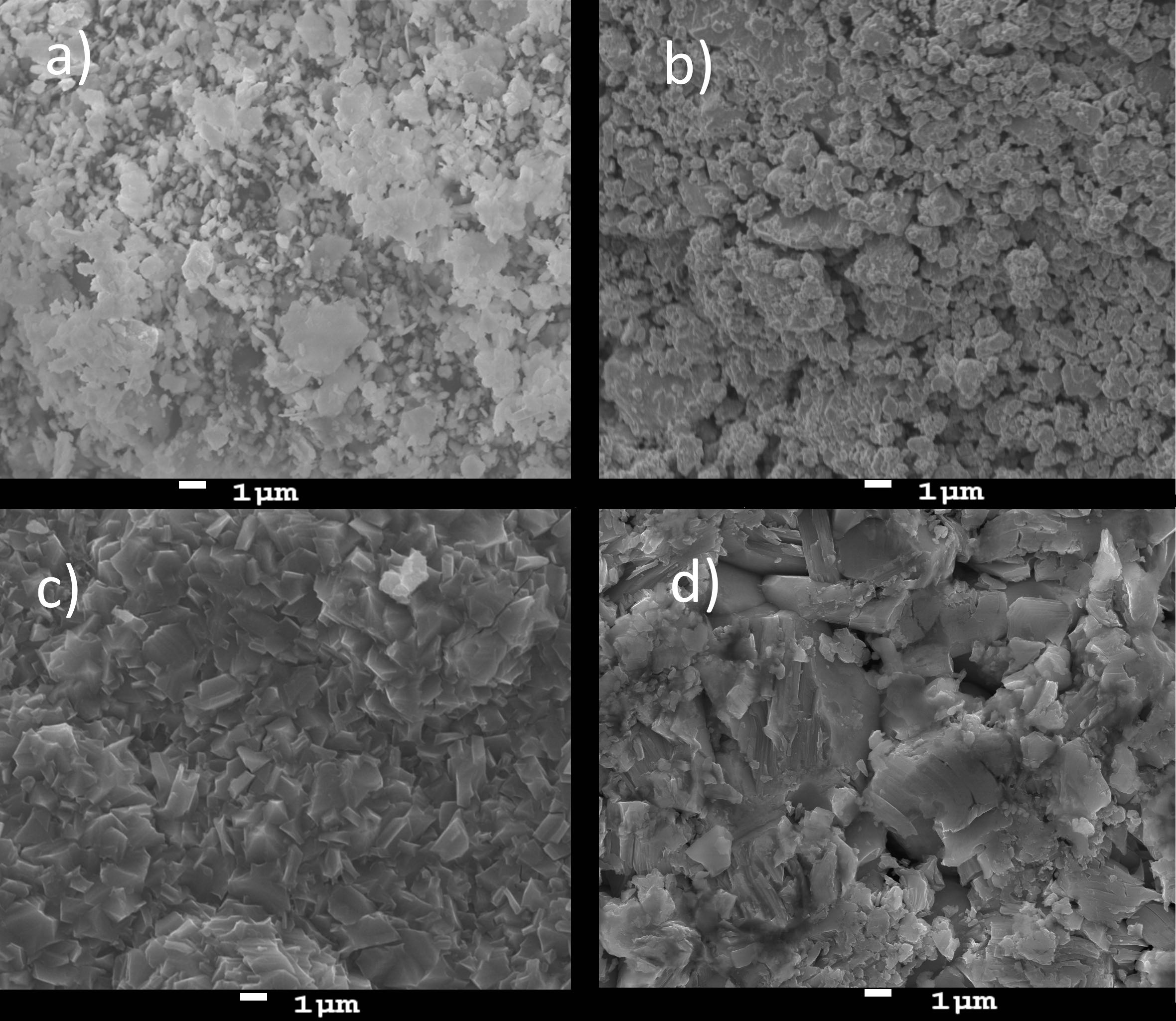}
	\caption{SEM images (x5000) of raw material after milling for (a) $BaFe_{1.9}Ni_{0.1}As_2$ and (b) $Ba_{0.6}K_{0.4}Fe_2As_2$, respectively. (c,d) The same samples heat-treated at $850^0C$ for 1 hour.  }
	
\end{figure}

\section{Superconducting Properties}
Superconducting properties were studied by AC susceptibility measurements down to 2K. The raw material after milling was thoroughly wrapped in PARAFILM obtained from Bemis in a glove box with an argon atmosphere to prevent interaction between the milled powder and the atmospheric oxygen and/or moisture; then, the material was placed into a sample holder for susceptibility measurements. The powder after MA shows no superconducting transition down to 2K in both cases (see Fig. 6), which confirms the formation of only an amorphous $Ba-122$ compound.  After annealing, the superconducting transition appears. Fig. 6 depicts the temperature dependence of the volume AC susceptibility measurements ($\chi'(T)$) for superconducting $BaFe_{2-x}Ni_{x}As_2$ and $Ba_{1-x}K_{x}Fe_2As_2$ samples after heat treatment at $850^0C$. One can observe a sharp superconducting transition emerging for all samples with various doping levels after annealing for 1 hour. A sharp transition and constant susceptibility value after the transition are evident to a high volume of the superconducting phase. The apparent shielding volume fraction of our samples was found to reach up to 85$\%$ at 4.2 K after demagnetisation correction \cite{Prozorov}. We determined the critical temperature values for near-optimum doped BFNA and BKFA compositions to be 18.5 K and 35 K, respectively, which are slightly below the optimally doped values \cite{Pervakov, Wang}, and coincide with the elemental analysis investigations of the MA materials by EDS. According to \cite{Pyon}, one should use the surplus of the dopant to obtain the exact doping level.

\begin{figure}[t]
	\includegraphics [width=1 \textwidth]{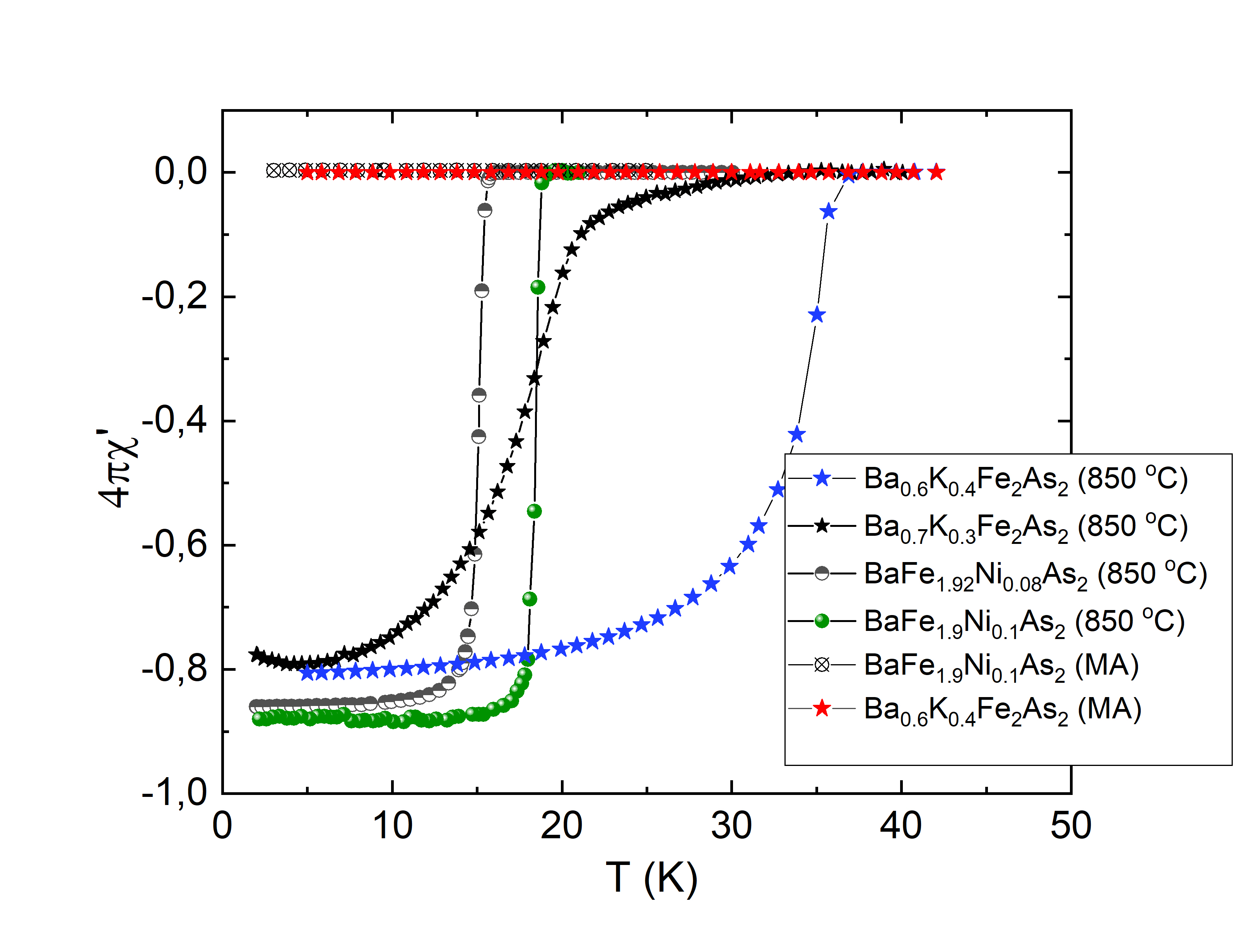}
	\caption {Temperature dependence of the real part ($\chi'$) AC susceptibility of BKFA and BFNA at zero DC magnetic field. The demagnetization effect has been taken into consideration \cite{Prozorov}.}
		
\end{figure}

\begin{figure}[t]
	\includegraphics [width=1 \textwidth]{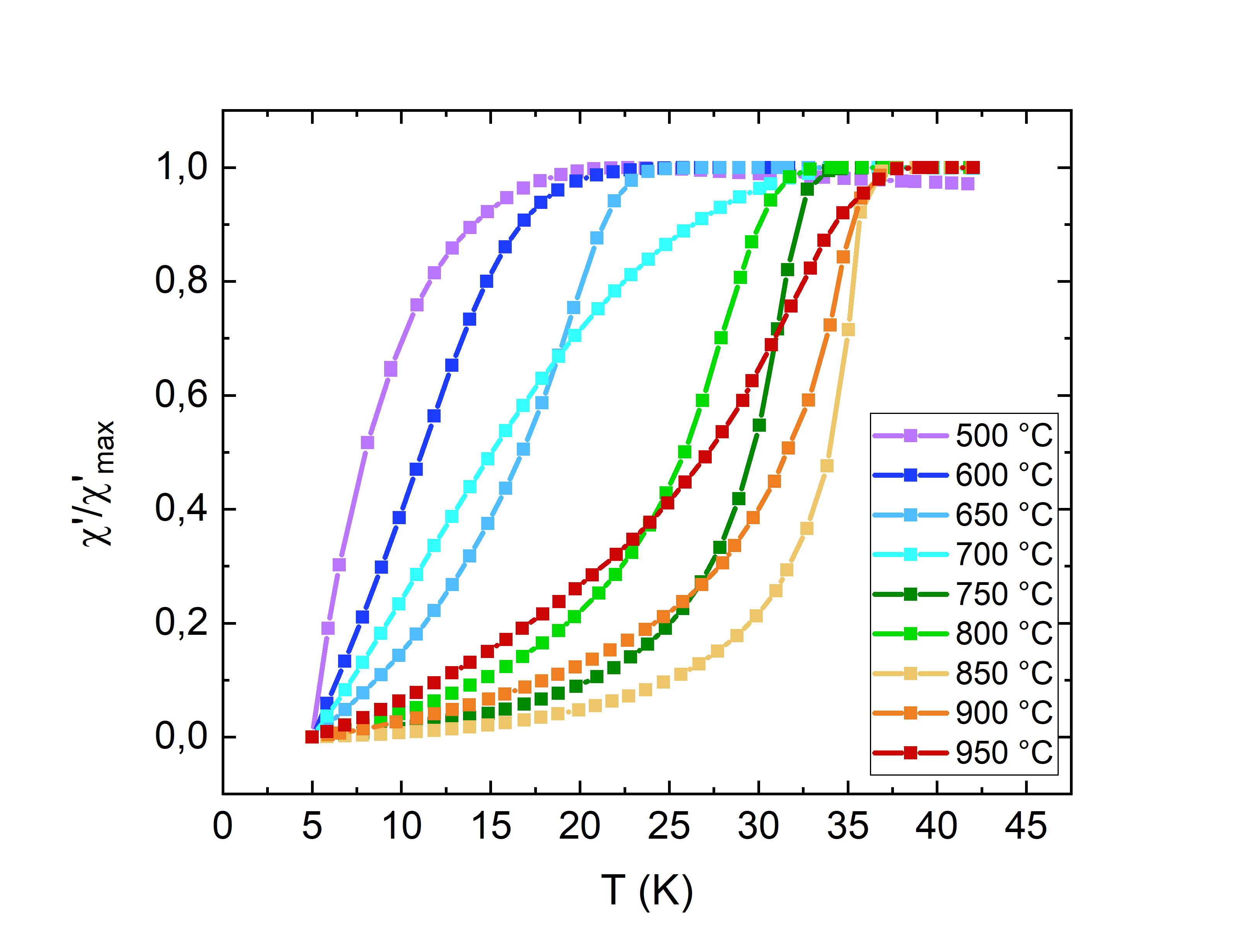}
	\caption{Temperature dependence of bulk $Ba_{0.6}K_{0.4}Fe_2As_2$ normalised real part ($\chi' /\chi'_{max}$) magnetic susceptibility after heat treatment at various temperatures for 1 hour.}
\end{figure}

As seen from Figure 7, maximum T$_c$ recorded for an annealing temperature of 850$^0C$. It coincides with the optimal temperature for $Sr-122$ \cite{Zang}. In the case of lower sintering temperatures, the samples show a small amount of superconducting phase and low $T_c$. Overshooting of the annealing temperature leads to a reduction of the superconducting phase volume and suppression of $T_c$. Thus, we found that an annealing temperature of approximately 850$^0C$ is optimal for various 122 superconductors for the formation of ceramic samples with high $T_c$.

\begin{figure}[t]
	\includegraphics [width=1 \textwidth]{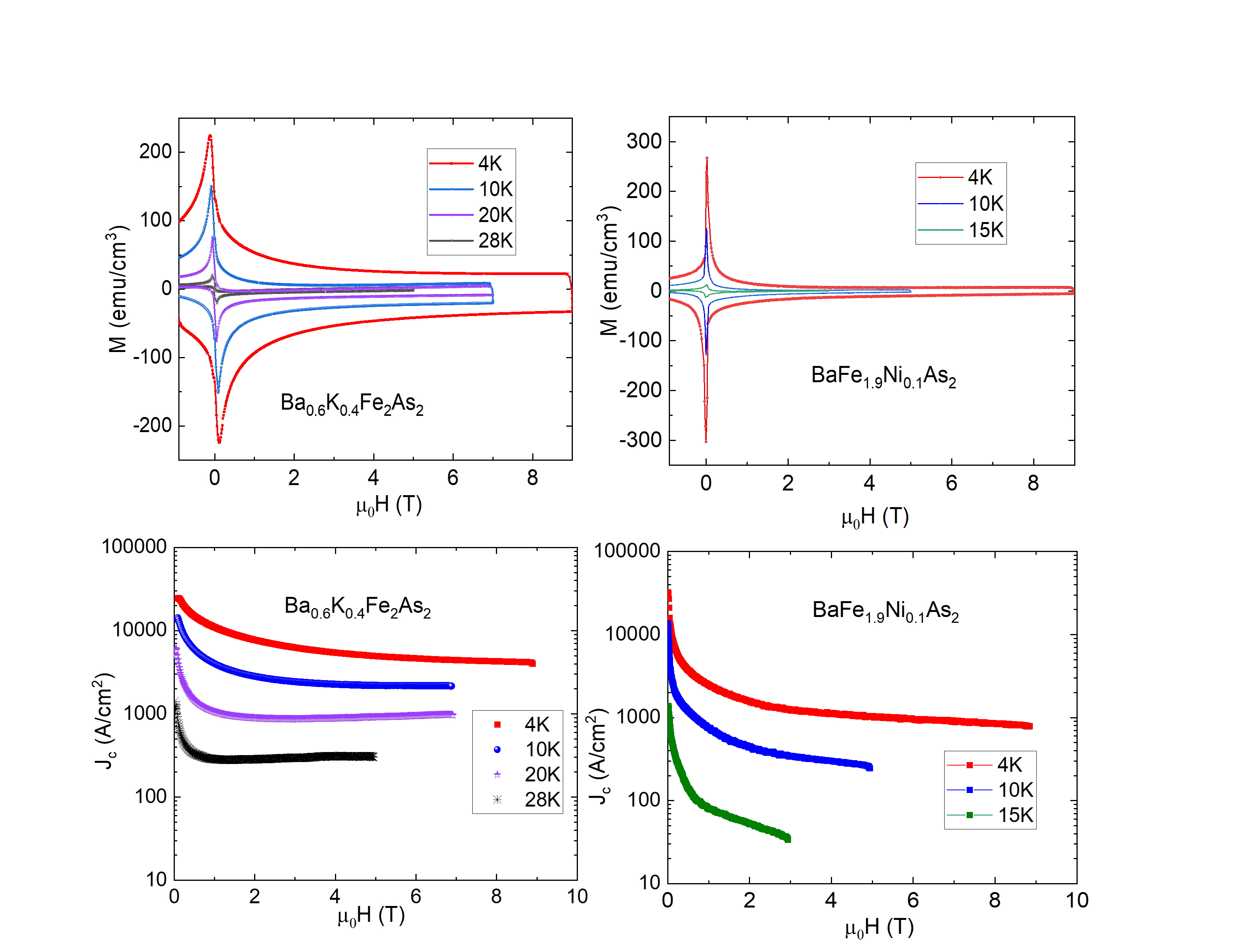}
	\caption{The isothermal magnetization loops, M(H), of $Ba_{0.6}K_{0.4}Fe_2As_2$ and $BaFe_{1.9}Ni_{0.1}As_2$. The critical current density ($J_c(H)$) calculated using the Bean critical state model \cite{Bean}, and plotted as a function of the magnetic field for $BaFe_{1.9}Ni_{0.1}As_2$ and $Ba_{0.6}K_{0.4}Fe_2As_2$, respectively.} 
\end{figure}

Figure 8 shows the M(H) loops for $BaFe_{1.9}Ni_{0.1}As_2$  and $Ba_{0.6}K_{0.4}Fe_2As_2$ samples in a magnetic field up to 9 T, respectively. The M(H) loops have a symmetrical shape that implies the domination of bulk pinning rather than pinning on the surface and geometrical barriers \cite{Abdel}. According to the Bean model \cite{Bean}, we estimate the critical current density ($J_c$) from the M(H) data, where  $J_c(H)  =  30 ×  \Delta M/D$,  with  D  being  the cylinder diameter and $\Delta M$ the width of the hysteresis loop. The $J_c$ value in zero field and T = 4K of $BaFe_{1.9}Ni_{0.1}As_2$ sample is approximately $31\times10^4 A/cm^2$ and for the $Ba_{0.6}K_{0.4}Fe_2As_2$ sample is $24\times10^4 A/cm^2$. Our $J_c$ values are comparable with the reported values of $J_c$ of other polycrystalline iron-based superconductors \cite{Shimoyama2}. 
	
Mechanical alloying, as it was mentioned before, is useful for pre-synthesis of IBS of the 11, 122, and 1111 systems. The following heat treatment takes much less time than the classical solid-phase reaction method. However, there are differences in the subsequent heat treatment for 11, 1111 and 122 superconductors. After mechanical activation, the 11 system bulks show maximum $T_c$ and superconducting phase volume after annealing at approximately 700$^0C$ for 12 hours in one or two stages \cite{Feng1,Liu}. Compounds of the 1111 system, according to \cite{Fang1,Fang2,Chen}, show a maximum critical temperature and the superconducting phase volume after annealing at 900-1000$^0C$ for 2-24 hours. However, the superconducting phase usually does not exceed 15-20$\%$. The most promising system for practical application is the 122 superconducting system, and its compounds were synthesised at temperatures of 750-900$^0C$ for 10-35 hours \cite{Li2,Pyon,Huang}. In our work, we found that the maximum amount of amorphous 122 phase is formed within 1-1.5 hours of MA. Following short-term annealing for 1 hour at an optimal temperature of approximately 850$^0C$, makes it possible to obtain polycrystalline material with a superconducting phase volume of up to 85$\%$. At the same time, long-term annealing, especially in the case of the alkaline doped 122 system, requires the compensation of arsenic and alkali metal loss.

\begin{table}[]
	\caption{\label{} Compilation of the BKFA and BFNA samples parameters. We show the initial formula, EDS, T$_c$ and cell parameters obtained for the investigated samples.}
	\renewcommand{\arraystretch}{1.3}
	\begin{center}	 
	\begin{tabular}{|c||c||c||c|c|}
		\hline
		Chemical Formula &  EDS  &T$_c$, K    & \multicolumn{2}{|c|}{Cell Parameter}\\
	
			\cline{4-5}
		     &   &  & a, \AA     &c, \AA  \\ 
		\hline 
		$BaFe_{1.9}Ni_{0.1}As_2$ & $Ba_{1.022}Fe_{1.907}Ni_{0.093}As_{2.003}$ &18.5K &3.9586(2) \AA &12.9820(7) \AA	\\ 
	\hline
		$BaFe_{1.92}Ni_{0.08}As_2$ & $Ba_{1.106}Fe_{1.921}Ni_{0.079}As_{1.776}$ &15K &- &-	\\ 
	\hline
		$Ba_{0.6}K_{0.4}Fe_2As_2$ & $Ba_{ 0.729}K_{0.271}Fe_{1.92}As_{1.92}$ &36K &3.9405(7) \AA &13.195(3) \AA	\\ 
	\hline
		$Ba_{0.7}K_{0.3}Fe_2As_2$ & $Ba_{0.734}K_{0.266}Fe_{2.15}As_{1.99}$ &25K &- &-	\\ 
	\hline

		\end{tabular}
	\end{center}
\end{table}

 \section{SUMMARY}
 The effect of ball-milling time on the structural and superconducting behaviour of $Ba_{1-x}K_{x}Fe_2As_2$ and $BaFe_{2-x}Ni_{x}As_2$ was studied. We show the formation of the amorphous non-superconducting phase in BKFA and BFNA compounds during mechanical alloying and determine the optimum milling time to be approximately 1-1.5h. 
 The compounds obtained for both samples exhibit superconductivity after annealing. The SEM images revealed the apparent formation of crystallites approximately 3-5$\mu$m in size, and magnetic susceptibility measurements show a sharp superconducting transition. The optimal short-range heat treatment temperature is found to be T = 850$^0C$. Mechanical alloying appears to be an up-and-coming technique for the large scale production of high-quality ceramic BKFA and BFNA superconductors.

\section{ACKNOWLEDGEMENT}

Work was done using equipment from the LPI Shared Facility Center and support by the Russian Foundation for Basic Research (RFBR project no. 17-29-10036).

\end{document}